\documentclass[10pt,prl,twocolumn,natbib,showkeys] {revtex4}

\usepackage[dvips]{graphicx}

\begin{document}

\title{Drastically suppressing the error of ballistic readout of qubits}
\author{Andrey L. Pankratov$^{1,2}$}
\email[]{alp@ipm.sci-nnov.ru}
\author{Anna V. Gordeeva$^{2,1}$}
\author{Leonid S. Kuzmin$^{2,3}$}
\affiliation{$^{1}$Institute for Physics of Microstructures of RAS, Nizhny Novgorod, 603950, Russia}
\affiliation{$^{2}$Laboratory of Cryogenic Nanoelectronics, Nizhny Novgorod State Technical University,
Nizhny Novgorod, Russia}
\affiliation{$^{3}$Chalmers University of Technology, Sweden}

\keywords{Long Josephson junctions, ballistic readout, thermal fluctuations, jitter.}

\begin{abstract}
The thermal jitter of transmission of magnetic flux quanta in long Josephson junctions is studied. While for large-to-critical damping and small values of bias current
the physically obvious dependence of the jitter versus length $\sigma\sim\sqrt{L}$ is confirmed, for small damping starting from the experimentally relevant $\alpha=0.03$ and below
strong deviation from $\sigma\sim\sqrt{L}$ is observed, up to nearly complete independence of the jitter versus length, which is exciting from fundamental point of view, but also intriguing
from the point of view of possible applications.
\end{abstract}

\maketitle

\begin{figure}[h]
\resizebox{1\columnwidth}{!}{
\includegraphics{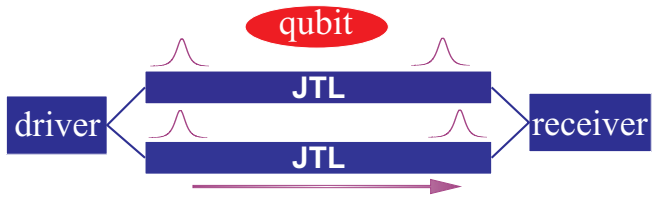}}
\caption{The illustration of the ballistic readout from the qubit using two Josephson transmission lines, where top one is
inductively coupled to the qubit.} \label{fig1}
\end{figure}

In the recent years superconducting circuits have attracted a considerable interest as promising devices for quantum computations\cite{q1,q2,q3,q4}.
The advantage of superconducting circuits in comparison with other types of qubits is the possibility to combine in one chip both
the qubits and the readout electronics. Mostly, dc SQUIDs are used for the readout from superconducting qubits. Recently \cite{SA} it has been suggested
to use for the readout the well-elaborated Rapid Single Flux Quantum devices (RSFQ), redesigned in a special way. The main idea of this type of readout
is the ballistic propagation of magnetic flux quanta along two separated Josephson transmission lines (JTL) \cite{SA}-\cite{HFSIS}, consisting of either
one long or series of short junctions having weak damping, see Fig. \ref{fig1}.
If one JTL is inductively coupled to the qubit, the magnetic field of qubit will change the speed of the fluxon (soliton), traveling from one JTL end
to another one, and thus will change the fluxon transmission time. This difference of fluxon transmission times in two JTLs can effectively be measured
using the existing RSFQ circuitry (SFQ receiver in Fig. 1), while the proper fluxons (SFQ pulses) can be produced in SFQ driver. The use of underdamped
JTLs together with the standard RSFQ circuitry in close vicinity to a qubit is very promising, since allows to eliminate expensive external pulse generators
and readout equipment, to nearly eliminate parasitic capacitances and inductances, altering the readout pulses, and to finally improve the readout fidelity
of the qubits.

It is obvious that increasing the JTL length, the precision of time difference measurement can be improved. However, different types of fluctuations restrict the readout
precision since they lead to the jitter of traveling fluxons. The importance to study jitter in Josephson junctions was first understood in \cite{Ryl}, see also
\cite{LRev}. Later jitter in short junctions has been studied both analytically and numerically \cite{PRL}-\cite{JAP}. However, as it has been understood and
experimentally verified \cite{Ter,Ter2}, for series of junctions the jitter increases as $\sigma\sim\sqrt{N}$, where $N$ is the number of junctions. Therefore, for JTLs
the jitter must increase as $\sigma\sim\sqrt{L}$, where $L$ is the JTL length, which also follows from simple theory presented in \cite{FSSK}. Since the fluxon traveling time $\tau$
is proportional to the JTL length ($\tau\sim L$), one must find a compromise length, leading a low readout error $\sigma/\tau$. However, to our knowledge, except simple reasoning
and the theory of \cite{FSSK}, the jitter in underdamped JTLs and long Josephson junctions have not been studied neither theoretically nor experimentally. The experimental
results devoted to RSFQ can not be directly applied since they are obtained for critically damped junctions rather than for underdamped. The present paper, therefore, is devoted
to detailed theoretical study of the jitter versus damping and the length in an underdamped long Josephson junction. While for large-to-critical damping and small values of bias current
we fully confirm $\sigma\sim\sqrt{L}$ dependence of the jitter versus length \cite{FSSK}, for small damping starting from the experimentally relevant $\alpha=0.03$ and below, we observe
strong deviation from $\sigma\sim\sqrt{L}$, up to nearly complete independence of the jitter versus length, which is exciting from fundamental point of view, but also intriguing
from the point of view of possible applications. In particular, this allows to drastically suppress the readout error by increasing the length of underdamped JTLs.

In the course of the present paper let us consider the long Josephson tunnel junction (JTJ) as a limiting
case of JTL with very little spacing between short junctions, much smaller than the Josephson penetration depth $\lambda_J$.
All calculations are performed in the frame of the sine-Gordon equation, giving a good qualitative description of
basic properties of long JTJ:
\begin{equation}
{\phi}_{tt}+\alpha{\phi}_{t}
-{\phi}_{xx}=i-\sin (\phi)+i_f(x,t),
\label{PSGE}
\end{equation}
where indices $t$ and $x$ denote temporal and spatial derivatives.
Space and time are normalized to the Josephson penetration length
$\lambda _{J}$ and to the inverse plasma frequency
$\omega_{p}^{-1}$, respectively, $\alpha={\omega_{p}}/{\omega_{c}}$
is the damping parameter, $\omega_p=\sqrt{2eI_c/\hbar C}$,
$\omega_{c}=2eI_cR_{N}/\hbar$, $I_c$ is the critical current, $C$ is
the JTJ capacitance, $R_N$ is the normal state resistance, $i$ is the dc overlap bias
current density, normalized to the critical current density $J_c$,
and $i_f(x,t)$ is the fluctuational current density. If the
critical current density is fixed and the fluctuations are treated
as white Gaussian noise with zero mean, its correlation function is:
$\left<i_f(x,t)i_f(x',t')\right>=2\alpha\gamma \delta
(x-x^{\prime})\delta (t-t^{\prime})$, where $\gamma = I_{T} /
(J_{c}\lambda_J)$ is the dimensionless noise intensity \cite{Fed},
$I_{T}=2ekT/\hbar$ is the thermal current, $e$ is the electron
charge, $\hbar$ is the Planck constant, $k$ is the Boltzmann
constant and $T$ is the temperature.

The boundary conditions that describe coupling to the environment,
have the form \cite{pskm}:
\begin{eqnarray}\label{x=0}
\phi(0,t)_{x}+r_L c_L\phi(0,t)_{xt}-c_L\phi(0,t)_{t t}=\Gamma, 
\\ \phi(L,t)_{x}+r_R c_R\phi(L,t)_{x t}+c_R\phi(L,t)_{t t}=\Gamma.
\label{x=L}
\end{eqnarray}
Here $\Gamma=0$ is the normalized magnetic field, and $L$ is the
dimensionless length of JTJ. The terms with the dimensionless
capacitances and resistances, $c_{L,R}$ and $r_{L,R}$, are the
RC-load of a JTJ placed at the left (input) and at the right
(output) ends, respectively.

For simplicity we assume that the bias current density is uniformly distributed along the space $i(x)=\rm{const}$.
As initial condition a kink $\phi(x,t)=4\arctan\left[\exp(x-x_0)\right]$ inside the junction is taken (for $x_0=5$).
Temporal and spatial intervals are chosen $\Delta t=\Delta x=0.01$, and it has been verified that further
decrease of the steps did not change the results. Two values of $RC$-load were tested: mismatched case
where simple boundary conditions $d\phi(x=0)/dx=d\phi(x=L)/dx=0$ were used, and perfectly matched case
$r_{L}=r_{R}=1$, $c_{L}=c_{R}=100$. Since the results are nearly the same, the curves for the
perfectly matched case (to suppress reflected waves from the junction ends, as required in experiments) are shown.

Two different definitions of the fluxon traveling time $\tau$ and the jitter $\sigma$ are used. One is the usual
mean first passage time of the boundary, where random time at which the soliton hits the right junction end is computed and
after the mean value and the standard deviation are calculated by averaging over 1000 realizations. Another definition
is based on the notion of the integral relaxation time \cite{ACP},\cite{PRL},\cite{Fed}: if the probability to find the soliton
inside the junction is computed $P(t)=\int_{0}^{L} W(x,t)d x$, the mean traveling time and the standard deviation (jitter) are
\begin{eqnarray}\label{prob}
\tau&=&\left< t \right >=\int_0^\infty t w(t)dt,\,\,\,
\sigma=\sqrt{\left< t^2 \right > - \left< t \right >^2}, \\
w(t)&=&\frac{\partial P(t)}{\partial t[P(\infty)-P(0)]}. \nonumber
\end{eqnarray}
Since we have investigated our task in the limit of small noise, both definitions give the same results within calculation precision as it must \cite{ACP}.

\begin{figure}[h]
\resizebox{1\columnwidth}{!}{
\includegraphics{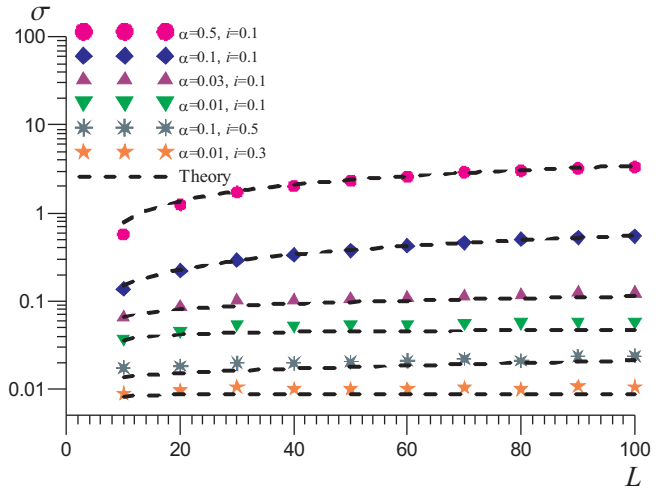}}
\caption{Jitter versus JTL length for various values of damping $\alpha$ and bias current $i$ for $\gamma=0.001$.
Symbols - simulations, dashed curves - theory (\ref{prob}),(\ref{P}).} \label{fig2}
\end{figure}
In Fig. \ref{fig2} the plots of the jitter versus JTL length are shown for various values of damping $\alpha$ and bias current $i$,
and for noise intensity $\gamma=0.001$. One can see that while for moderate values of damping $\alpha=0.5$ and $\alpha=0.1$
the curves are perfectly fitted by the dependence $\sigma\sim\sqrt{L}$, starting from $\alpha=0.03$ the jitter is proportional to
$\sigma\sim\sqrt[4]{L}$ and for decreasing values of damping and increasing bias current the jitter tends to a constant, which is
the main and quite surprising result of the present paper.

\begin{figure}[h]
\resizebox{1\columnwidth}{!}{
\includegraphics{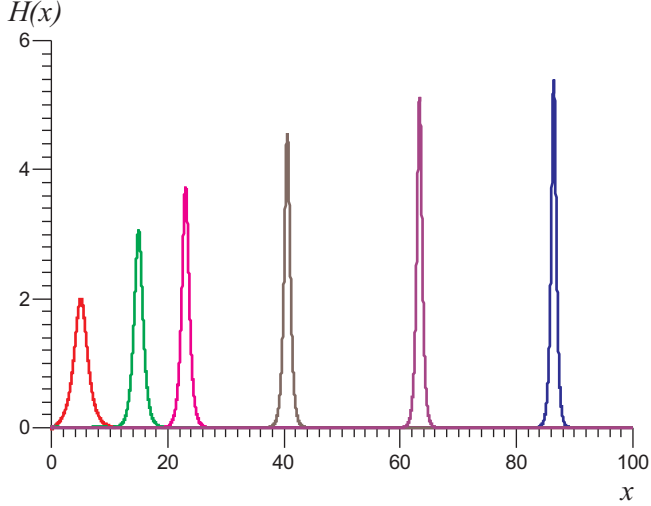}}
\caption{Snapshots of the soliton, accelerating and contracting during the motion along JTL, $i=0.1$ and $\alpha=0.03$. } \label{sol}
\end{figure}
In order to understand the origin of the observed result let us consider the evolution of soliton shape. From Fig. \ref{sol} one can see the snapshots of the fluxon accelerating and contracting during its motion along JTL. Analyzing the soliton dynamics for different values of damping and bias current it has been noted that if the damping is moderate, and the bias current is small, then the soliton's shape weakly changes with time. In this case the dependence of the jitter is well described by $\sigma\sim\sqrt{L}$.
However, if the damping is small, see, e.g., Fig. \ref{sol} for $i=0.1$ and $\alpha=0.03$, the Lorentz contraction is observed, the soliton width seriously decrease
near the output boundary in comparison with the input one. Considering this as the main explanation that can be given for violation of $\sigma\sim\sqrt{L}$ dependence of the jitter,
we have tried to obtain more advanced theory, since the formula of \cite{FSSK} does not take into account the nonstationarity of the fluxon dynamics.

\begin{figure}[h]
\resizebox{1\columnwidth}{!}{
\includegraphics{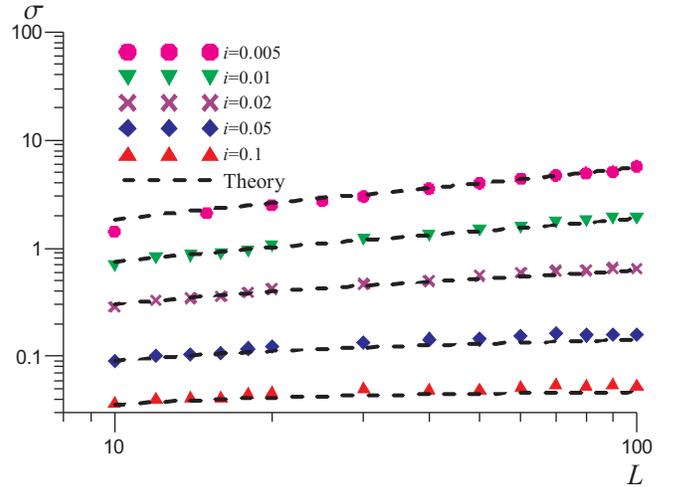}}
\caption{Jitter versus JTL length for different values of bias current $i$ and $\alpha=0.01$, $\gamma=0.001$. Symbols - simulations, dashed curves - theory  (\ref{prob}),(\ref{P}). The dependence changes from $\sqrt{L}$, (top curve, circles) to $\sqrt[4]{L}$ (diamonds).} \label{figi}
\end{figure}

\begin{figure}[h]
\resizebox{1\columnwidth}{!}{
\includegraphics{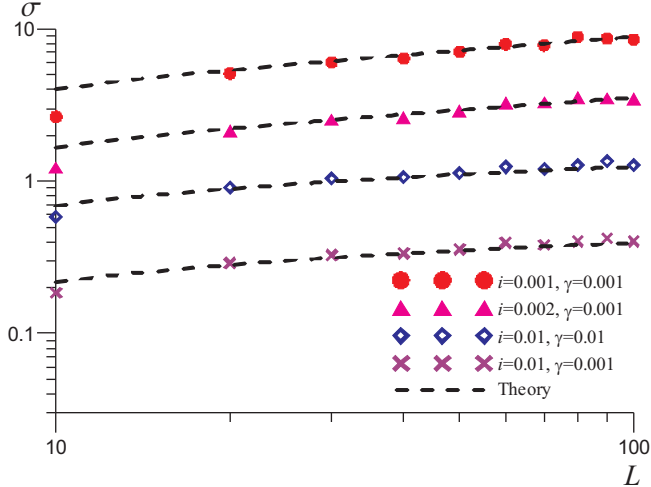}}
\caption{Jitter versus JTL length for very low damping $\alpha=0.001$. Symbols - simulations, dashed curves - theory (\ref{prob}),(\ref{P}). All curves are well fitted by $\sqrt[4]{L}$ dependence.} \label{lowdamp}
\end{figure}

\begin{figure}[h]
\resizebox{1\columnwidth}{!}{
\includegraphics{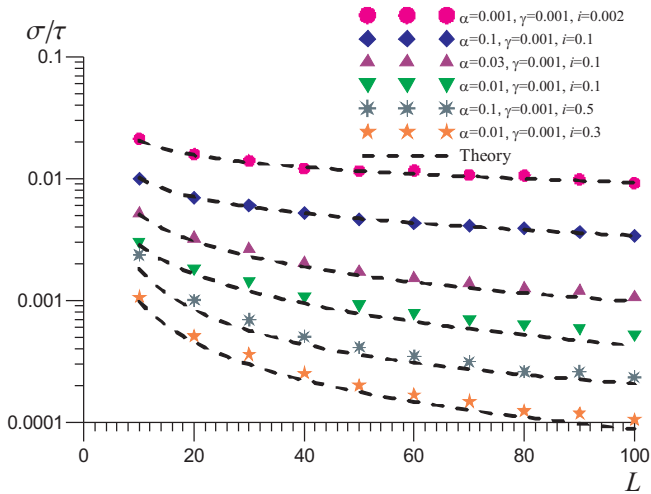}}
\caption{The readout error $\sigma/\tau$ versus JTL length. Symbols - simulations, dashed curves - theory (\ref{prob}),(\ref{P}). The dependence changes from $1/\sqrt{L}$ (top) to $1/L$ (bottom).} \label{st}
\end{figure}

Starting from Eq. (\ref{PSGE}), one can write the equation for the momentum $p(t)=-\frac{1}{8}\int_{-\infty}^{+\infty}\phi_x\phi_tdx$ of the center of the soliton \cite{McS} with account of effect of noise \cite{shlw}
\begin{equation}\label{mom}
\frac{dp}{dt}=-\alpha p+\frac{1}{4}\pi i+\xi(t),
\end{equation}
where $p=v/\sqrt{1-v^2}$, $v(t)$ is the fluxon velocity, and the noise intensity $\left<\xi(t)\xi(t')\right>=\frac{\alpha\gamma}{4\sqrt{1-v^2}}\delta(t-t')$ with $\gamma$ the same as in Eq. (\ref{PSGE}). Directly from Eq. (\ref{mom}), neglecting the effect of noise, one can derive the expression for the fluxon velocity $v(t)$ for $v(t=0)=0$
\begin{equation}\label{v}
v(t)=\frac{1}{\sqrt{1+\frac{1}{\left[\frac{\pi i}{4\alpha}(1-e^{-\alpha t})\right]^2}}}.
\end{equation}
From Eq. (\ref{mom}) one can derive the following equation for the location of the center of the fluxon $X(t)$
\begin{equation}\label{X}
\ddot{X}(t)=-\alpha \dot{X}(t) (1-v^2)+(\frac{1}{4}\pi i+\xi(t))\sqrt{(1-v^2)^3},
\end{equation}
where $\ddot{X}(t)$ and $\dot{X}(t)$ stand for temporal derivatives. This equation is too complex to be solved analytically, and we will use the following approximations. First, let us neglect random component of velocity (which can be done in the limit of small noise), so $v(t)$ is described by Eq. (\ref{v}). Second, let us also neglect by the term $(1-v^2)$ in front of the first temporal derivative $\dot{X}(t)$, so Eq. (\ref{X}) transforms into the equation of massive Brownian particle, but with the noise intensity, depending on time. This Brownian diffusion is described by the Gaussian probability density, since the noise $\xi(t)$ is Gaussian. Therefore, one can use the approach described in the chapters 1 and 3 of the book \cite{Lan}, and can get the following expression for the variance $D(t)$ of the process $X(t)$, taking into account the nonstationarity of the noise intensity
\begin{equation}\label{D}
D(t)=\frac{\gamma}{4\alpha}\int_0^t\left[1-2e^{-\alpha y}+e^{-2\alpha y}\right](1-v(y))^{5/2}dy,
\end{equation}
where $v(y)$ is given by Eq. (\ref{v}).
Analogically, the mean $m(t)$ of $X(t)$ can be derived (with $X(t=0)=X_0$)
\begin{equation}\label{m}
m(t)=X_0+\int_0^t v(y) dy.
\end{equation}
Substituting the mean and the variance into the Gaussian probability distribution one can obtain the probability to find the soliton inside the junction
\begin{equation}\label{P}
P(t)=1-{\rm erfc}\left[(L-m(t))/\sqrt{2D(t)}\right]/2.
\end{equation}
Substituting $P(t)$ into the definitions (\ref{prob}), the mean traveling time $\tau$ and the standard deviation $\sigma$ can be computed.
The comparison between computer simulation results and the theory is presented in Fig. \ref{fig2},\ref{figi}-\ref{st}. One can see good agreement for larger damping, while for smaller damping the theory slightly underestimates the numerical results, presumably due to our simplifications in Eq. (\ref{X}). Nevertheless, the agreement is good enough and correctly describes the qualitative behavior, confirming violation of $\sqrt{L}$ dependence of the jitter.

Let us perform calculations for fixed small value of damping $\alpha=0.01$
and for various bias current values, see Fig. \ref{figi}. It is seen that for small values of bias current the dependence $\sigma\sim\sqrt{L}$ is maintained,
while for larger values starting from $i=0.01$ the deviation is observed. It should be noted, however, that one must think about thermal budget of
both qubits and its readout electronics, trying to decrease the total heat, since usually cryostats have rather low thermal power
at low temperatures. Nevertheless, the bias current values of order $i=0.01$, at which the deviation from the dependence $\sigma\sim\sqrt{L}$ is observed, seem to be rather low to maintain low heat emission.

It is known that with decreasing the temperature of JTL from 4K to 50 mK the damping decreases by one order of magnitude \cite{FSSK}.
This allows to operate with much smaller bias currents to decrease heat emission, see Fig. \ref{lowdamp}. The curves, calculated for $\alpha=0.001$, demonstrate good fitting
to the dependence $\sigma\sim\sqrt[4]{L}$ even for very low bias current $i=0.001$.

Finally, let us consider the ratio of jitter to the traveling time $\sigma/\tau$, which determines the error of measurement (readout error) due to thermal fluctuations. In Fig. \ref{st}
the plots of $\sigma/\tau$ are presented versus JTL length, for different parameters $\alpha$ and $i$. One can see that contrary to the conclusions of Ref. \cite{FSSK} larger
bias current leads to smaller readout error $\sigma/\tau$.

In conclusion, the thermal jitter of transmission of magnetic flux quanta in long Josephson junctions is studied. While for large-to-critical damping and small values of bias current
the known dependence of the jitter versus length $\sigma\sim\sqrt{L}$ is confirmed, for smaller damping the strong deviation from $\sigma\sim\sqrt{L}$ is observed, up to nearly complete independence of the jitter versus length, which is exciting from fundamental point of view, but also intriguing
from the point of view of possible applications.

The work is supported by RFBR project 09-02-
00491, Dynasty Foundation, Human Capital Foundation, and by the
Act 220 of Russian Government (project 25).

\end{document}